# A Generalized Web Component for Domain-Independent Smart Assistants

Yusuf Sermet

*Electrical and Computer Engineering*

*University of Iowa*

Iowa City, IA, USA

muhammedyusuf-sermet@uiowa.edu

Ibrahim Demir

*Civil and Environmental Engineering*

*University of Iowa*

Iowa City, IA, USA

Ibrahim-demir@uiowa.edu

**Abstract**

This article introduces an open-source web component, Instant Expert, which allows robust and efficient integration of a natural language question answering system to web-based platforms in any domain. Web Components are a set of web technologies to allow the creation of reusable, customizable, and encapsulated HTML elements. The Instant Expert web component consists of the user input (i.e. text, voice, multi-selection), question processing, and user interface modules. Two use cases are developed to demonstrate the component's features, benefits, and usage. The goal of this project is to pave the way for next-generation information systems by mitigating the challenges of developing voice-enabled and domain-informed smart assistants for communicating knowledge in any domain.

**Keywords:** *smart assistants; knowledge generation; intelligent systems; web components; shadow DOM*

## 1. Motivation and Significance

Immense amount of data is constantly being generated in a variety of fields including environmental, biological, and physical sciences due to the rapid advancements in monitoring and computational techniques [1-2]. The massive data is in need of efficient tools and intermediates for its management, analysis, visualization, and communication [3]. Web-based information systems (IS) serve as one-stop platforms to access, analyze, and explore information effectively for decision-making purposes [4-5]. Although current systems are proved to be successful to communicate and analyze data efficiently, they still suffer from the complexity and the higher learning curves due to the limitations of conventional interaction methods. Users of information systems (e.g. public, workers, managers, decision makers, organizational leaders) often look for a certain piece of knowledge for which they may have to master the functionalities and resources provided by the IS. This is especially tedious and discouraging for users who are not continuous visitors to the system.

To free the users from the nuances and complications of ISs, human-like interactions using natural language (NL) is vital. Next generation information systems are expected to act as automated domain experts which are required to have the ability to comprehend domain-specific NL queries, determine the intent and the parameters of the query, apply human-like reasoning to generate the direct answer out of curated and structured data, and present it in natural language [6]. Examples of such systems in the commercial setting include Apple Siri, Google Assistant, and Amazon Alexa [7]. The major contributions of augmenting information systems with domain-focused smart assistants are ease of use, providing accessibility support via voice recognition, and allowing the integration of the system to a variety of communication channels.

In this project, we have developed Instant Expert, an open-source web component that can serve as a boilerplate to build and integrate voice-enabled smart assistants for web-based information systems. Web components provide a standard to encapsulate a web application into a single reusable component by utilizing several modern web technologies [8]. One of the most profound advantages of web components is that they allow the reuse of complex web applications, which may require expertise to develop, by simply importing the JavaScript file that defines the web component with a line of code. Another profitable aspect of them is that they can be designed in a way to prevent potential conflicts (e.g. JavaScript variables, CSS rules, HTML attributes) with the hosting web site [9]. Prior examples of web components in the literature manifest its potential and benefits [10-14].

Our web component allows users to ask natural language questions supporting text input, voice input, and selection from a predefined list of questions. It retrieves the answer to the user's question by making an HTTP (Hypertext Transfer Protocol) request to the provided knowledge engine. Instant Expert is built using advanced web technologies including Web Components with Shadow DOM (Document Object Model) and CSS (Cascading Style Sheets) Containment as well as recent capabilities such as the Web Speech API (Application Programming Interface). The component can be used in a variety of contexts and domains while protecting the integrity and performance of the component and the hosting platform. The component has been implemented, deployed, and tested as part of 2 use cases (i.e. The Flood Expert, Microsoft Project Answer Search) to demonstrate its benefits and serve as a guide for its adoption.

The presented component will make it possible for any information system on any domain to have its own voice-enabled smart assistant to instantly provide factual responses to complex queries. It can grow the system's visibility and increase user retention and satisfaction due to providing the user with the information they desire without a hassle. The component can especially be valuable for individual developers, academic research groups, and small companies that may not have the resources for the development of smart assistants for their organization. The increasing availability of free and paid services to build question-answering engines and NLP tools makes it viable and economically feasible to develop the core of such systems for a variety of use cases ranging from simple websites to complex information systems. The Instant Expert provides a generic framework to offer standardized, robust, and efficient smart assistants by freeing the developers from dependence to any service in the backend.

## 2. *Software Description*

### 2.1. *Software Architecture*

The Instant Expert web component consists of three major components (Figure 1). The input layer is responsible for handling list, text and voice inputs. Question processing module combines the communication protocols and methods with the knowledge engine, which will serve as the natural language question answering system. User interface (UI) layer represents the component's visual elements as well as managing the user interaction. The web component is implemented using HTML5 (Hypertext Markup Language), JavaScript, and CSS and depends on the jQuery library. The usage of web frameworks (e.g. Polymer, Stencil) is considered when designing the web component, though consciously avoided to minimize the learning curve and eliminate the dependency on any framework's abstraction.

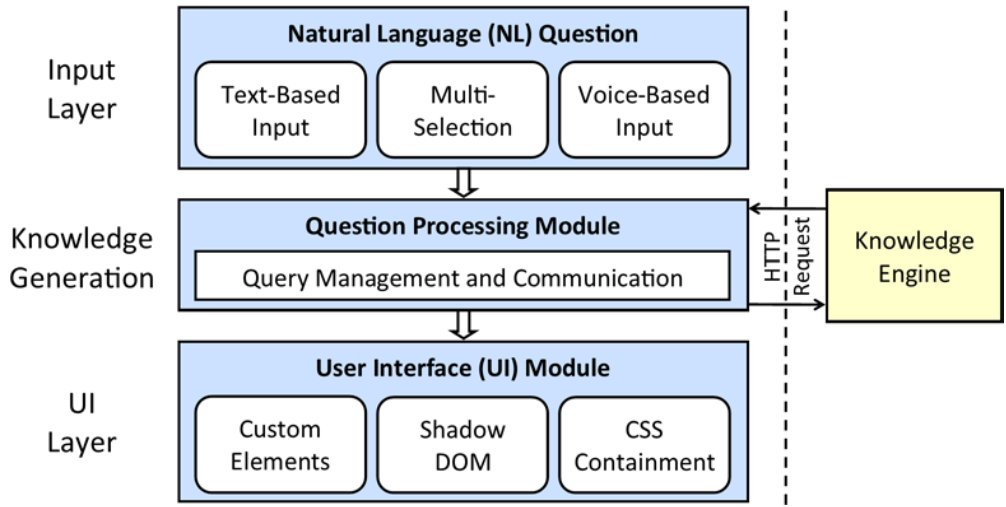

Figure 1: Instant Expert System Architecture. Yellow box represents external services outside of Instant Expert's scope.

## 2.2. Software Functionalities

### 2.2.1. Web Component

The Instant Expert is built upon the features provided by the Web Components. Web Components are a collection of web technologies combined with the purpose of creating reusable, customizable, and encapsulated HTML elements [15]. It is mainly powered by three web technologies. Custom Elements enables the creation of new HTML elements. HTML templates offer the mechanism to define HTML content that can be instantiated during runtime instead of getting rendered when the page is loaded. And, Shadow DOM provides encapsulation of an element's features with a shadow tree associated with the web component [16]. The major consideration when designing the component is to prevent it to affect the visual and functional integrity of the hosting web site, as well as to prevent it from getting affected by the hosting web site. Shadow DOM prevents potential integration complications by creating a shadow root under the custom HTML element and rendering it separately from the main document DOM. These new technologies are not yet widely adopted in both the industry and especially in academia despite the remarkable benefits they bring. Another aim of this paper is to serve as an example to Web Components' advantages by reducing the technicality.

The front end of the component consists of a text box with two buttons for enabling list selection and voice recognition (Figure 2). The web component is activated (i.e. initialized) via a button placed at the middle-left of the window to toggle the component's visibility. The design is kept at a minimum to allow maximum customization of the component to blend into the hosting web site naturally. All default visuals of the component are changeable by simply providing the visuals' absolute or relative path.

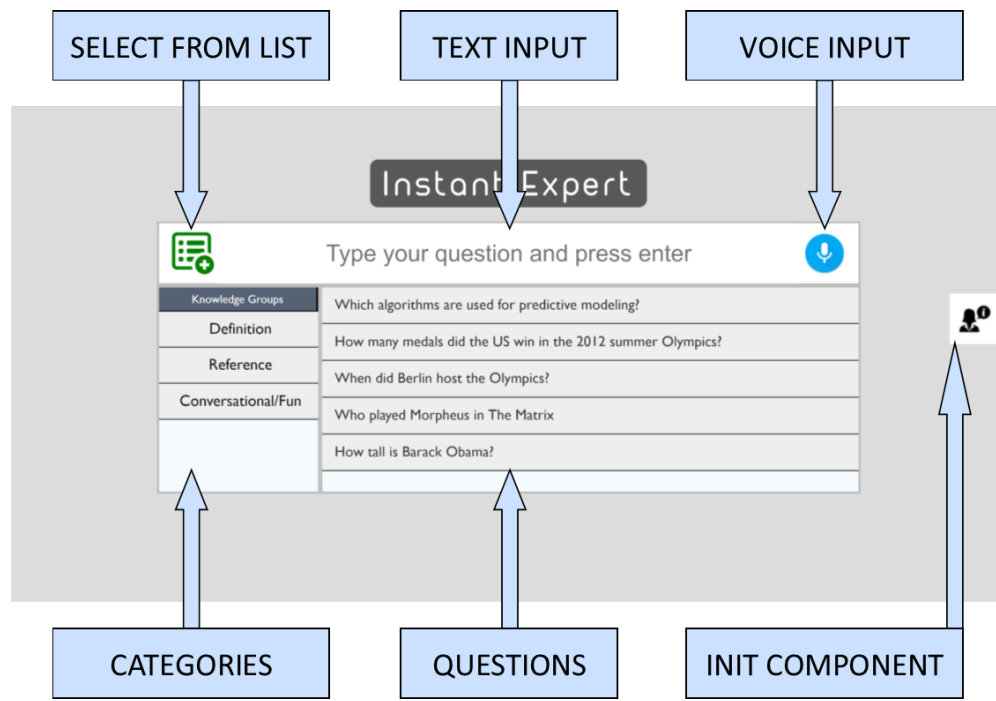

Figure 2: User input options for Instant Expert

*2.2.2. CSS Containment*

Containment is a CSS feature that aims to isolate a contained element's contents from the rest of the document, as much as possible. Main purpose of the containment is to provide optimization and offer stability in rendering and painting of web pages by providing a standardized way. Browsers are always looking to make optimizations when rendering a page. Containment provides a standardized way to tell the browser where and how it can optimize without breaking the intended functionality. When using third-party DOM, such as the Instant Expert, containment can prove to be useful to sandbox the component to protect and increase the performance of the page. It should be noted that containment is not a security feature and is not aimed to provide a full encapsulation.

There are 4 main types of containment which can either be used individually or in groups (Figure 3). The details of each containment type can be found in the specification document published by the World Wide Web Consortium (W3C) [17]. For the purposes of Instant Expert, the Content Containment has been used which combines Layout, Paint, and Style Containments. A web component already brings the containing functionality; however, Content Containment introduces significant performance benefits and decreases rendering runtime. This is especially valuable to protect the performance of the website that integrates the Instant Expert. As of June 2019, CSS Containment is supported by default by the latest versions of major browsers (i.e. Google Chrome, Opera, Mozilla Firefox Nightly, and Microsoft Edge). Another advantage of using Content Containment is that it will not affect the functionality of Instant Expert even if the client browser does not support it.

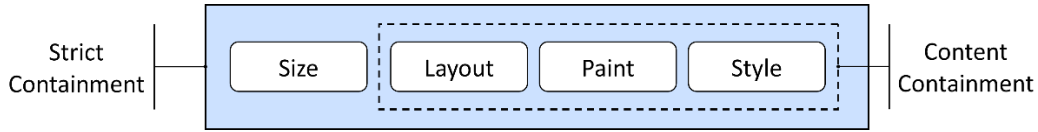

Figure 3: Types of CSS Containment

### 2.2.3. User Input

There are 3 ways for a user to interact with the system; a) manually typing the question to a text box, b) invoking voice recognition to ask the question using a microphone, and c) selecting from a predefined list of questions (Figure 2). Text input is the most common type of interaction due to the search engine culture. Having a predefined list of questions allow the user to explore the system and better understand its capabilities. Voice-enabled communication is supported using Web Speech API, which is an experimental technology that defines a JavaScript API to integrate speech recognition and speech synthesis functionality into web pages. As of June 2019, the speech recognition is supported by the latest versions of Google Chrome, Opera, and Microsoft Edge browsers which constitute approximately the 57.8% of all users in the world. Speech synthesis is supported in all major browsers (e.g. Google Chrome, Mozilla Firefox, Opera, Microsoft Edge, and Safari). The component checks the client browser at initialization to test if Web Speech API is supported and disables speech features if not supported. The component allows the incorporation of third-party speech recognition and synthesis APIs, however, requires modification of the component's source code. Upon the construction of the natural language question in text format, all input types eventually follow through the same flow to be passed to the engine.

### 2.2.4. Question Processing

The web component retrieves the answer for the input question by making an HTTP POST request to a natural language question answering engine using the webhook link provided with the 'engine' attribute of the element. The only parameter passed to the engine is the question text using the parameter key 'question'. This key can be changed by setting the attribute 'engineDataKey'. The component expects a response from the engine in JSON format with a key-value pair where the key is 'resultText' and value is the natural language answer. This key can be changed by setting the attribute 'engineResponseKey'. The request times out if the response is not received in 2 seconds. If the engine and the web page that integrates the Instant Expert are not hosted from the same origin, then the engine should be able to handle requests from origins outside of its own by setting up Cross-Origin Resource Sharing (CORS).

## 2.3. Software Adaptation

There are a couple of steps to successfully adapt and implement the presented component as a smart assistant. The specific actions are detailed in the GitHub repository with demonstrative examples. Table 1 summarizes the main required actions that need to be taken for initialization as well as their purposes.

**Table 1.** Action items to initialize and deploy a smart assistant to any website using Instant Expert

| Action Item | Subtask | Scope and Description |
|---|---|---|
| Setup Dependencies | Import jQuery | In-class methods of the component benefit the abstractions provided by jQuery when making HTTP requests. |
| Setup Instant Expert | Import Instant Expert | Import the source code of the web component as a script. |
| Use Instant Expert | Add HTML element | Add the HTML element <instant-expert> into the code where Instant Expert is desired. |
| | Set engine webhook | Set the element's 'engine' attribute to the webhook link to make POST requests to retrieve answer to user's question. |

According to the web standards, a page should be served on a secure connection if HTML5's Speech Recognition API is used. Thus, the webpage that the component is built-in and the webhook link for the engine must provide *https* connections if the voice-input is desired. The component has all types of inputs enabled by default, including text and voice inputs as well as a selection from a list. The component enables and encourages its customization in accordance with the needs of the application. Table 2 summarizes the attributes and properties to customize the Instant Expert web component.

**Table 2.** Attributes and properties to customize the Instant Expert

| Name | Expected Value | Default Value | Scope and Description |
|---|---|---|---|
| **Attributes** | | | |
| engine | *String* | *Unset* | The webhook link that will be used to make POST requests to get the responses to user queries. |
| engineDataKey | *String* | *'question'* | The parameter name to pass the user's question in the POST request. |
| engineResponseKey | *String* | *'resultText'* | The parameter name to access the answer in JSON object returned by the engine. |
| logo-src | *String* | *Base64 Image* | The image path for the component's logo. |
| logo-hidden | *Boolean* | *True* | Toggle logo's visibility: True=visible, false=hidden |
| textbox-placeholder | *String* | *'Type your question and press enter.'* | The placeholder text that will be displayed on the input text box. |

| no-question-list | *Boolean* | *False* | Disable the question list: True=disabled |
| no-voice | *Boolean* | *False* | Disable the voice input: True=disabled |
| expert-button-src | *String* | *Base64 Image* | The image path for the component's logo. |
| **Properties** | | | |
| setQuestions | *Question List* | *N/A* | Set the example questions for selection from a list. As an input parameter, it takes an array of duplets: each consisting of the question text and its category. |

### *3. Illustrative Examples*

#### *3.1. The Flood Expert*

The presented component has been implemented in the field of flooding as part of the Iowa Flood Information System (IFIS; http://ifis.iowafloodcenter.org). The IFIS is a one-stop web-based platform for real-time and historical flood-related data management, analysis, and visualization including flood inundation maps, flood conditions and forecasts [18]. The component uses the Flood AI [19] as its engine which powers its ontological background [20], question models, data resources, and capabilities using the IFIS for flood-related knowledge generation. Figure 4 shows the component's integration with the IFIS to establish a precedent. The Flood Expert is developed at the Iowa Flood Center and has not been made available with this project source code due to its substantial data and resource dependence.

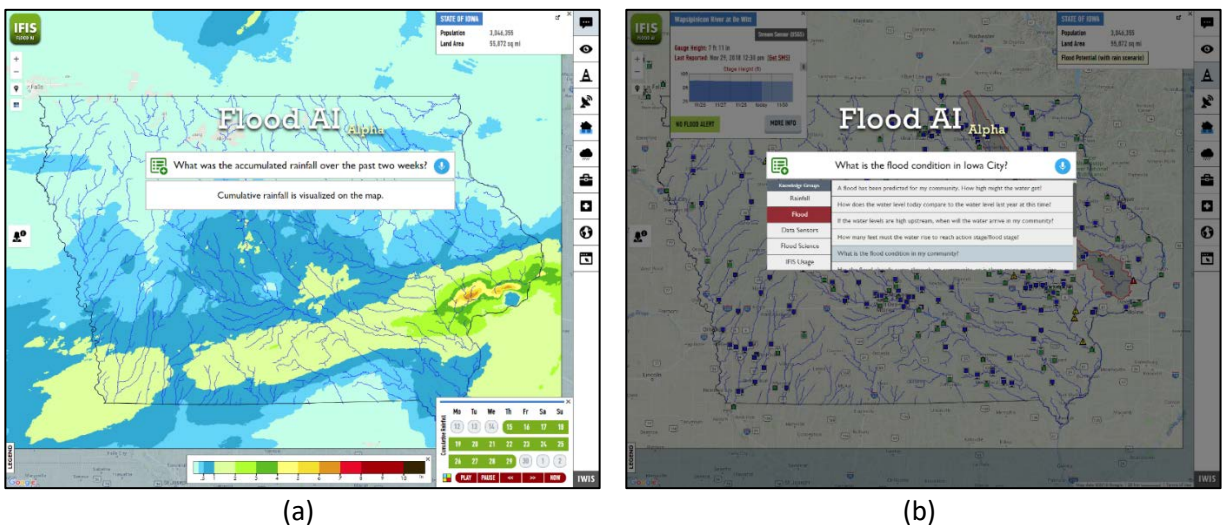

| (a) | (b) |

Figure 4: (a) Usage of Flood Expert to control the IFIS functionalities via natural language questions. (b) The list interface that allows the exploration of the supported questions grouped into broad categories.

#### *3.2. Microsoft Cognitive Labs – Project Answer Search*

The second use case of the component is the development of a generic question answering assistant using the Project Answer Search by Microsoft Cognitive Labs. Project Answer Search is an experimental technology to instantly answer natural language user queries with factual responses [21]. This use case is specifically developed to be a complete solution that (a) will serve as an adoption guide to the users, (b) can be easily accessed to try the presented software, (c) and can be conveniently reproduced for production use or ensure the component's correctness. There are two parts constituting this use case; developing a question-answering engine as the backend and developing a website that implements the Instant Expert component integrated with the engine. The backend is implemented as a Node.js application and served on a cloud platform (i.e. Heroku). The source code for both the backend and the website are available on Instant Expert GitHub repository in the examples directory along with the directions necessary for reproduction. Figure 5 shows this use case in action.

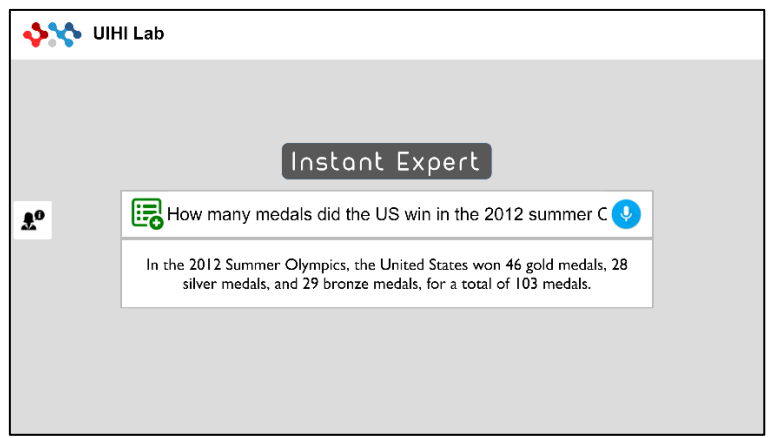

Figure 5. Usage of Microsoft's Project Answer Search with Instant Expert web component.

## 4. *Impact*

- The Instant Expert web component provides the front end, interaction capabilities, and user input mechanisms for a generalized, customizable, and domain-independent natural language question answering mechanism for web-based information systems.
- It utilizes a variety of advanced web technologies including Web Components with Shadow DOM and CSS Containment to provide an isolated, robust, and efficient solution, and frees the developers from the complications of integrating third-party components into existing web platforms.
- Many of the web technologies that are utilized in this project are still not yet widely adopted in both the industry and especially in academia. The Instant Expert serves as a guide and demonstration of these technologies' benefits while reducing technicality.
- The web component decreases the complexity and the amount of work needed to implement a smart assistant and makes it possible for developers, research groups and companies in any domain to enrich their web platforms and information systems.

## 5. *Conclusions*

This project is aimed at providing a generalized open-source web component (i.e. The Instant Expert) to create smart assistants for next-generation information systems. The component is built to work with open-source and free services to allow users to create the minimum viable product without any financial investment. In case the demands and the scope of the assistant grow, the component allows the customization of its services to scale with the emerging needs. For future work, Instant Expert's scope and capabilities can be expanded into virtual reality (e.g. A-Frame), and augmented and virtual reality (i.e. HoloLens and Samsung Gear VR). The preliminary work for integrating Instant Expert into VR is reported in [22].

## Acknowledgements

This project is based upon work supported by the Iowa Flood Center and the University of Iowa.

**Metadata**

*Table 1 – Code metadata*

| Nr | Code metadata description | *Metadata* |
|---|---|---|
| C1 | Current code version | *v1.0* |
| C2 | Permanent link to code/repository used of this code version | *https://github.com/uihilab/instant-expert* |
| C3 | Legal Code License | *MIT License* |
| C4 | Code versioning system used | *Git* |
| C5 | Software code languages, tools, and services used | *HTML, JavaScript, CSS* |
| C6 | Compilation requirements, operating environments & dependencies | *Any internet browser that supports Web Components, jQuery* |
| C7 | If available Link to developer documentation/manual | *https://github.com/uihilab/instant-expert/blob/master/README.md* |
| C8 | Support email for questions | *muhammedyusuf-sermet@uiowa.edu* |